\documentclass[12pt]{article}
\usepackage[tablesfirst,notablist,nomarkers]{endfloat}
\usepackage{amssymb,amsbsy,amsmath,amsfonts,amssymb,amscd}
\usepackage{epsf}
\newcommand{\T}[1]{{\mathbf{#1}}}   
\newcommand{\TT}[1]{{\mathcal{#1}}} 
\newcommand{\ie}{{\em i.e., \/}}       
\DeclareTextSymbol{\degre}{OT1}{23}
\begin{document}

\title{Nomarski imaging interferometry to measure the displacement field of MEMS}

\author{Fabien Amiot, Jean Paul Roger}

\maketitle 

\begin{abstract} We propose to use a Nomarski imaging
interferometer to measure the out-of-plane displacement field of
MEMS. It is shown that the measured optical phase arises both from
height and slope gradients. Using four integrating buckets a more
efficient approach to unwrap the measured phase is presented, thus
making the method well suited for highly curved objects. Slope and
height effects are then decoupled by expanding the displacement
field on a functions basis, and the inverse transformation is
applied to get a displacement field from a measure of the optical
phase map change with a mechanical loading. A measurement
reproducibility of about 10 pm is achieved, and typical results
are shown on a microcantilever under thermal actuation, thereby
proving the ability of such a set-up to provide a reliable
full-field kinematic measurement without surface modification.
\end{abstract}


\section{Introduction}

\par The increasing interest for micro-electro-mechanical systems
(MEMS), especially when they are used as micro-mechanical sensors
\cite{lavrik}, leads one to focus on the mechanical behavior of
micro-objects. First, the standardized mechanical tests at the
macro scale have been adapted to the micro one, assuming an
homogeneous stress or strain state. Bending \cite{weihs88} and
tensile tests \cite{read93,haque03}, as well as fatigue or creep
tests \cite{mems_handbook} are performed since years, providing a
global kinematic response of the tested object.
\par However, classical photolithography processes use visible
light to transfer a mask onto a wafer surface. Then, light
diffraction limits the achievable accuracy of the geometric shape
of the resulting micro-objects, and the dimensions margins tends
to increase compared to the dimensions themselves as the object's
size decreases. The homogeneous stress (or strain) assumption
usually satisfied when performing mechanical tests at the
macro-scale is no longer reasonable, and one has to deal with
heterogeneous mechanical tests. As a consequence, one has to
perform a spatially resolved kinematic measurement instead of a
global one.
\par Moreover, as their size decreases, the surface on volume ratio significantly increases, and the behavior of
micro-objects tends to be dominated by surface effects instead of
volume ones. As a consequence, measuring the displacement field
should avoid any surface modification or any contact. In addition,
the surface roughness of MEMS is usually very low, so that
measurement techniques involving surface-generated speckle
\cite{hung79} may be difficult to implement \cite{aswendt03}.
\par An optical interferometric imaging set-up is then well suited to
measure displacement fields. The polarization interferometer
\cite{francon} proposed herein is derived from the one initially
proposed by Nomarski \cite{nomarski}. Dealing with surface
topographies, it has been previously used to determine the mean
slope of tilted samples \cite{lessor79} or to get an image of the
roughness of polished surfaces using a multichannel Nomarski
microscope \cite{gleyzes94,gleyzes95,gleyzes97}.
\par After recalling the relation between the optical phase induced by the sample-Wollaston prism group and the measured
intensity, the twofold origin of the optical phase is detailed. A
shot-noise limited detection is described, and the inversion
method developed to convert an optical phase change into an
out-of-plane displacement is presented. An example is finally
provided on the measurement of the displacement field of a
thermally loaded micro-cantilever.

\section{Measuring a differential topography}
\subsection{Experimental set-up}\label{ss:nom}
\begin{figure}[h]
        \centerline{\epsfxsize=.5\hsize \epsffile{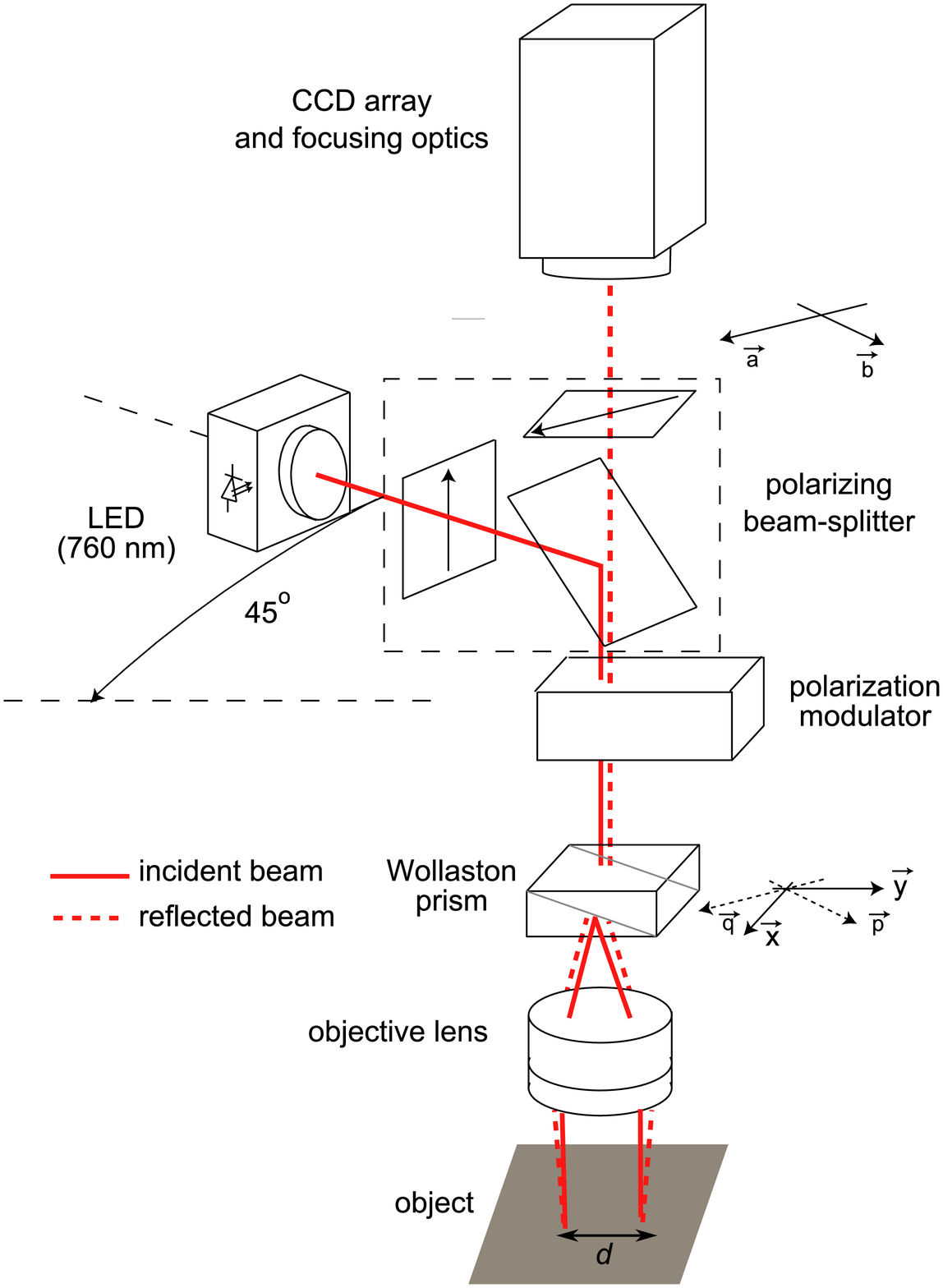}}
        \caption{Schematic view of the basic interferential microscopy imaging set-up.}
        \label{fig:nomarski}
\end{figure}
The basic interferential microscopy imaging set-up is shown in
Fig.~\ref{fig:nomarski}. A spatially incoherent light source
(LED,~$\lambda =760$~nm) illuminates a polarization beam-splitter.
The beam reflected by the beam-splitter goes through a
polarization modulator and is initially polarized at $45 \degre$
of the axes of a Wollaston prism. This splits the beam into two
orthogonally polarized beams at a small angle between each other.
These beams are focused upon the sample by an objective lens.
After reflection and recombination by the Wollaston prism, the
beam goes through the polarization modulator and the polarization
beam-splitter. The transmitted beam is finally focused on a CCD
array. The polarization beam-splitter behaves as crossed linear
polarizers mounted at $45 \degre$ of the axes of the Wollaston
prism and of the polarization modulator. One finally gets the
interference between two images of the sample, shifted by the
Wollaston prism of the distance $d$. The resulting interference
pattern is recorded on a CCD array (DALSA CA-D1, $ 256 \times 256$
pixels, 8 bits).

\subsection{Interference pattern obtained with a Nomarski
imaging interferometer} Let us denote $\T{p}$
(Fig.\ref{fig:nomarski}) the polarisation direction of the beam
incident on the Wollaston prism. The Wollaston prism
shear-direction is denoted $\T{y}$. The vector $\T{a}$ denotes the
polarisation direction of the beam impinging on the CCD array. The
orthogonal directions, in the prism's plane, are denoted $\T{q}$,
$\T{x}$, $\T{b}$, respectively. One uses a light emitting diode,
so that we denote $E_0$ the amplitude of the non polarized wave
impinging on the polarizing beam-splitter. If $t_p$ is its
amplitude transmission factor, and $\epsilon_p$ the attenuation
factor for the (ideally) suppressed component, the Wollaston prism
is illuminated by two orthogonally polarized beams, which electric
fields are $E_{pp}\T{p}$ and $E_{pq}\T{q}$, with
\begin{eqnarray}
E_{pp}=E_0 t_p \\
E_{pq}=E_0 \epsilon_p t_p
\end{eqnarray}
For a non polarized light source (LED), these two beams are
incoherent, and should be treated separately. $\epsilon_p t_p$ is
the transmission factor in the stop direction, so that
$\epsilon_p=0$ is the perfect polarizer case. For each beam, the
Wollaston prism splits the beam into two orthogonally polarized
beams (\ie a $\T{x}$ and a $\T{y}$ component), and the light goes
through the path (objective - sample - objective - Wollaston
prism). Let us consider that this results in a phase shift $\phi$
between the $\T{x}$ and the $\T{y}$ components. Let us denote
$t_n$ the transmission factor of the path (Wollaston prism -
objective - sample - objective - Wollaston prism), $t_a$ the
analyser's transmission factor, and $\epsilon_a t_a$ its further
attenuation factor in the stop direction. The total intensity $I$
impinging locally on the CCD array is the sum of the intensity
arising from the two incoherent beams
\begin{equation}
I = I_p + I_q
\end{equation}
Assuming that $(\T{x},\T{p})=(\T{x},\T{b})=\frac{\pi}{4}$, one
obtains \cite{lessor79}
\begin{eqnarray}\label{eq:form_inter}
I&=&I_0+A \cos (\phi) \\
I_0&=&\frac{|E_0 t_a t_n
t_p|^2}{2}(1+\epsilon_a^2)(1+\epsilon_p^2)\\
A&=&\frac{|E_0 t_a t_n t_p|^2}{2}(1-\epsilon_a^2)(1-\epsilon_p^2)
\end{eqnarray}
The contrast of the interference pattern
\begin{equation}
\frac{A}{I_0}=\frac{(1-\epsilon_a^2)(1-\epsilon_p^2)}{(1+\epsilon_a^2)(1+\epsilon_p^2)}
\end{equation}
equals $1$ for perfect polarizers ($\epsilon_a=\epsilon_p=0$), and
decreases when $\epsilon_a$ or $\epsilon_p$ increase. Adding an
error on the orientation of the polarization beam-splitter leads
to the same expression, so that the equation (\ref{eq:form_inter})
is considered general enough to describe real interference
patterns.
\begin{figure}[htb]
       \centerline{\epsfxsize=.5\hsize \epsffile{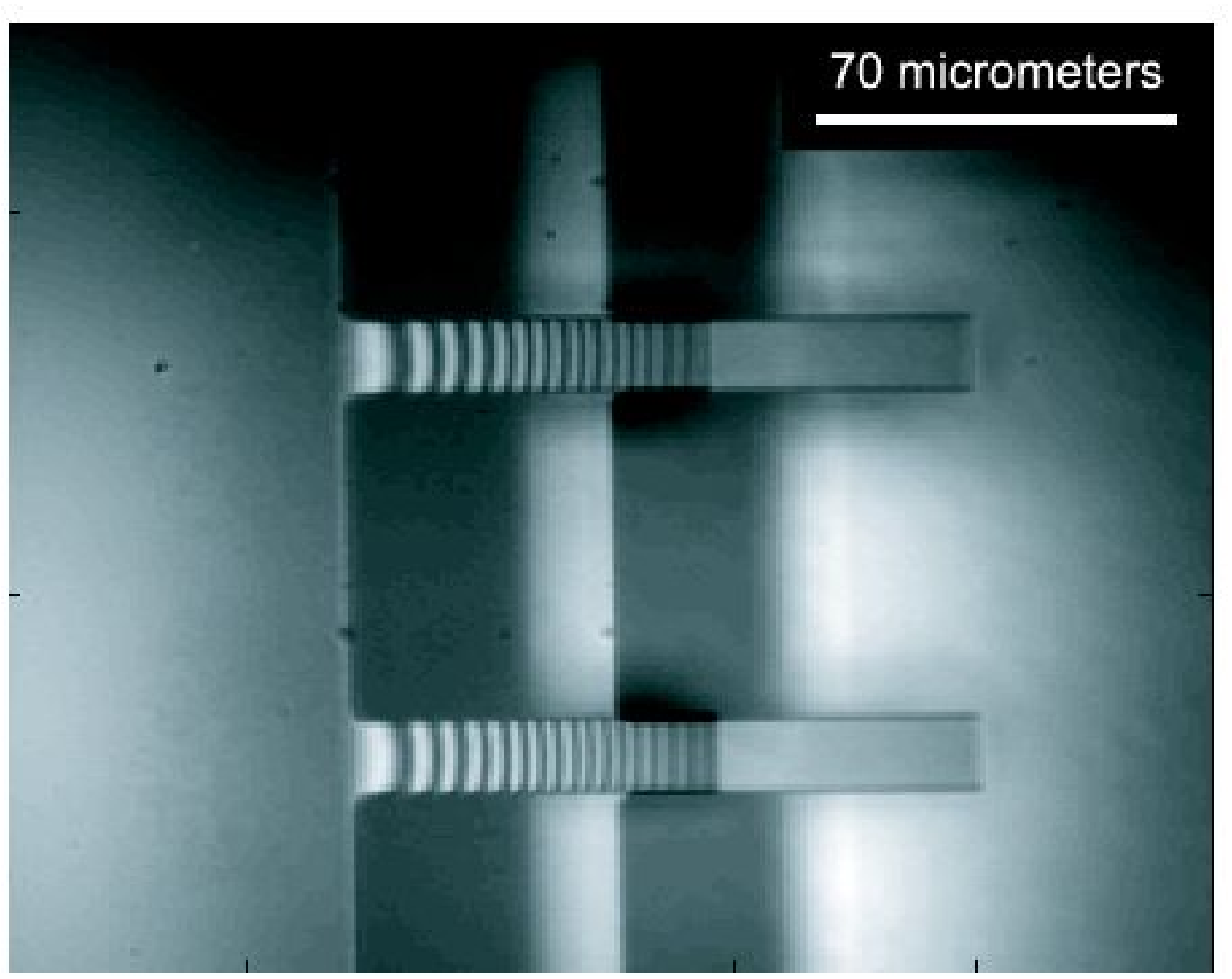}}
       \caption{Typical interference pattern obtained in water with two $70 \times 20 \times 0.84 \mu m^3$ microcantilevers and a shear distance $d \simeq 50 \mu m$ (NA=0.3).}
       \label{fig:ima_interf}
\end{figure}
A typical interference pattern obtained in water with two 70
micrometers long micro-cantilevers and a shear distance $d \simeq
50 \mu m$ is shown on figure \ref{fig:ima_interf}. The two sheared
images are clearly distinguishable. The optical phase range covers
almost 15 interference fringes. The closely packed interference
fringes, as well as the quite short correlation length of the used
light source (almost $15 \mu m$) reduce the contrast of the
interference pattern, thereby limiting the quantitative use of
obtained phase map. Let us assume that the optical phase $\phi$
introduced by the path (Wollaston prism - objective - sample -
objective - Wollaston prism) may be decomposed in a term $\phi_0$
arising from the Wollaston prism and a contribution $\phi_m$
arising from the object
\begin{equation}\label{eq:partition_phi}
\phi=\phi_0+\phi_m
\end{equation}
The section \ref{ss:phi_wollaston} exhibits the phase directly
arising from the Wollaston prism, whereas the section
\ref{ss:phi_topo} exhibits the phase arising from the topography
of the sample.

\subsection{Optical phase arising from the Wollaston prism}

\label{ss:phi_wollaston}

\subsubsection{Optical path functions}

\begin{figure}[htb]
       \centerline{\epsfxsize=.5\hsize \epsffile{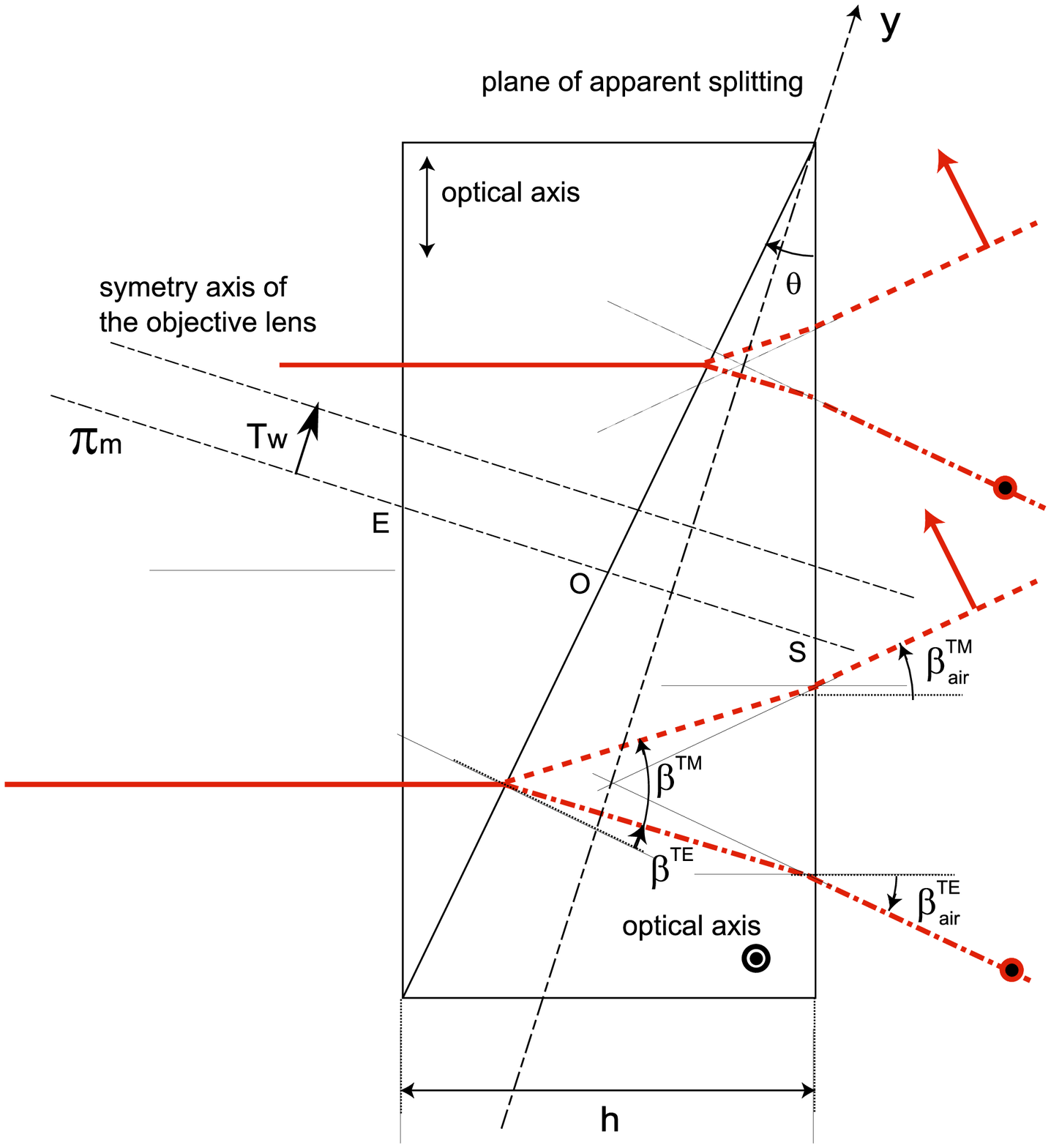}}
       \caption{Schematic view of a Wollaston prism.}
       \label{fig:angles_wollaston}
\end{figure}
Let us consider for simplicity an ``ideal'' Wollaston prism, which
geometry is described in the plane defined by both the optical
axis of the system and the $y$ direction on figure
\ref{fig:angles_wollaston}. Fig. \ref{fig:angles_wollaston} shows
the decomposition of a ray impinging orthogonally on the prism
into two emerging rays and the angles definition. Snell-Descartes'
laws at the interface between the two half-prisms read
\begin{eqnarray}
n_E \sin (\beta^{TM})&=&n_o \sin (\theta) \\
n_o \sin (\beta^{TE})&=&n_E \sin (\theta )
\end{eqnarray}
where $n_o$ is the ordinary refractive index of the used
birefringent material, and $n_E$ is the extraordinary one.
Assuming an ambiant media with a refractive index equals to 1,
Snell-Descartes' laws for the exit face of the prism read
\begin{eqnarray}
\sin (\beta^{TM}_{air})&=&n_E \sin (\beta^{TM}-\theta ) \\
\sin (\beta^{TE}_{air})&=&n_o \sin (\beta^{TE} - \theta)
\end{eqnarray}
If all the angles are small
\begin{equation}
\beta^{TM}_{air} - \beta^{TE}_{air} \simeq 2 (n_o-n_E) \theta
\end{equation}
The emerging rays appear to split on a plane, called plane of
apparent splitting (PAS), which is here located inside the prism.
The use of a modified Wollaston prism \cite{montarou99} allows one
to move the PAS outside the prism. This plane is considered to be
perpendicular to the figure's plane. Its position is one of the
fundamental characteristics of the prism. Investigating the
imaging properties of the system, let us denote $\alpha_1$ the
angle of an impinging ray with respect to the normal of the
entrance interface in the figure's plane. Considering any
dependance on $\alpha_1$ is then moving in the field of view along
the shear direction. The optical path travelled by the $TE$ (resp.
$TM$) polarized ray through the prism $l^{TE}(y,\alpha_1)$ (resp.
$l^{TM}(y,\alpha_1)$) when the apparent splitting occurs at
position $y$ on the PAS (the origin will be defined later) depends
on $\alpha_1$. Assuming that all the angles are small,
\begin{eqnarray}
l^{TE}(y,\alpha_1) &\simeq& n_{E}\left(\frac{h}{2}+y \theta
\right)+n_{o} \left(\frac{h}{2}-y \theta \right) + \frac{h \alpha_1 n_1 \theta}{2}\left(\frac{n_o}{n_E}+\frac{n_E}{n_o} \right) \nonumber \\
l^{TM}(y,\alpha_1) &\simeq& n_{E} \left(\frac{h}{2}-y \theta
\right)+n_{o}\left(\frac{h}{2}+y \theta \right) + \frac{h \alpha_1
n_1 \theta}{2}\left(\frac{n_E}{n_o} \left( 2- \frac{n_E}{n_o}
\right) +1 \right)
\end{eqnarray}
These two functions will be referred as optical path functions.

\subsubsection{Optical phase when the PAS matches the objective rear focal plane}

\begin{figure}[htb]
       \centerline{\epsfxsize=.5\hsize \epsffile{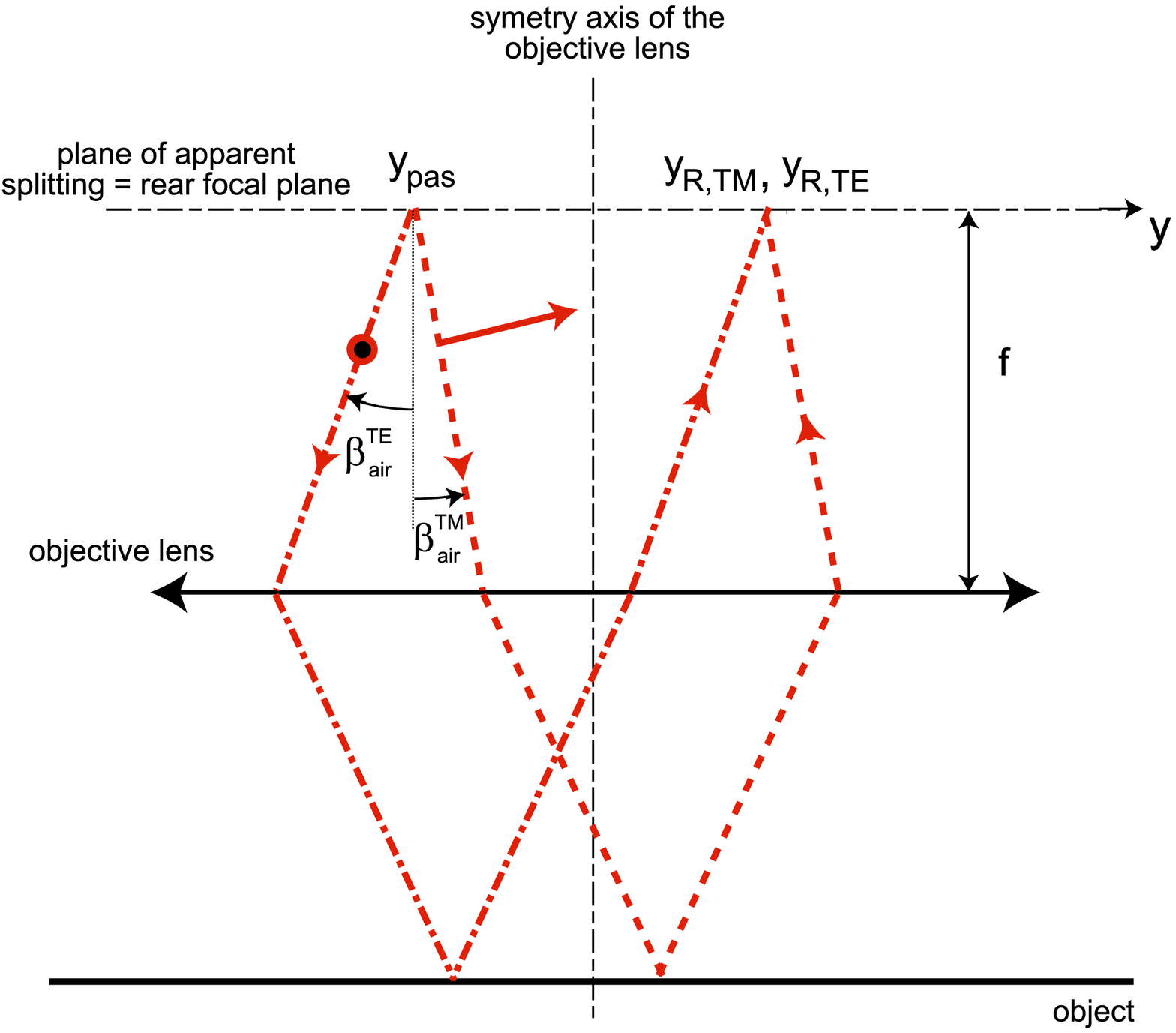}}
       \caption{Ray tracing for a plane object.}
       \label{fig:focal_wollaston0}
\end{figure}
Assuming that the PAS of the prism matches the objective rear
focal plane, Fig. \ref{fig:focal_wollaston0} presents the ray
tracing for the two emerging rays of Fig.
\ref{fig:angles_wollaston}. These two emerging rays intersecting
the PAS at the $y_{pas}$ cross the PAS, after reflection on a
plane object, at the the same point $y_{R,TE}=y_{R,TM}$ when the
object is orthogonal to the lens axis. This last point is
symmetric of the first with respect to the lens axis. Let us thus
define the median plane of the prism $\pi_m$ as orthogonal to both
the figure's plane and the PAS, passing by the $O$ point (see Fig.
\ref{fig:angles_wollaston}). This point is defined to satisfy the
condition $EO=OS$. Let us then consider that the distance between
the $\pi_m$ plane and the symmetry axis of the lens is $T_w$. Any
point of the PAS may be described either by the abscissa $y$ in
the prism's frame (the origin is at the intersection with $\pi_m$)
or by the abscissa $\tilde{y}=y-T_w$ (relative to the symmetry
axis of the lens). In the prism's frame, the rays emerge from the
PAS at $y_{pas}=T_{w}-\tilde{y_t}$, and the reflected rays go
through the PAS at $y_{R,TE}=y_{R,TM}=T_{w}+\tilde{y_t}$. The
optical paths $l_{ar}^{TE}$ and $l_{ar}^{TM}$ travelled to the
sample and back are deduced from $l^{TE}(y,\alpha_1)$ and
$l^{TM}(y,\alpha_1)$ and allow one to compute the optical phase
\begin{equation}\label{eq:delta_trans}
\phi_{W0}=\frac{2 \pi}{\lambda}(l_{ar}^{TE}-l_{ar}^{TM})= \frac{2
\pi}{\lambda} \theta \left( 4(n_{E}-n_{o})T_{w} + h \alpha_1 n_1
\left( \frac{n_o}{n_E} + \frac{n_E}{n_o} (\frac{n_E}{n_o}-1) -1
\right) \right)
\end{equation}
The first term in (\ref{eq:delta_trans}) no longer depends on
$y_t$ but on the position of the prism with respect to the
symmetry axis of the lens $T_{w}$, and is homogeneous in the field
of view. If the birefringence of the used material is denoted
$n_E-n_o=n_o \epsilon$, the additional optical phase difference
(second term) proportional to the incidence angle $\alpha_1$ is
shown to grow as $\epsilon^2$ whereas the first term scales as
$\epsilon$, so that the added optical phase $\phi_{W0}$ may be
considered homogeneous in the field of view. For a quartz-made
Wollaston, $\epsilon \simeq 10^{-2}$ so that the added optical
phase may be considered homogeneous in the field of view. This is
no longer true when the PAS doesn't match the rear focal plane of
the objective, because the reflected rays no longer cross the PAS
at the same point (see Fig. \ref{fig:focal_wollaston0}). Assuming
that the PAS remains parallel to the rear focal plane, at a
distance $\delta_{PAS}$ and using the same method, the additional
optical phase difference $\phi_{PAS}$ is
\begin{equation}
\phi_{PAS} \simeq \frac{8 \pi}{\lambda} \theta \alpha_1
\delta_{PAS} (n_E-n_o)
\end{equation}
thus introducing a linear phase along the Wollaston shear
direction.

\subsection{Optical phase arising from the object
topography}\label{ss:phi_topo}
\subsubsection{Optical phase arising from height variations}\label{sss:phase_marche}

\begin{figure}[htb]
       \centerline{\epsfxsize=.5\hsize \epsffile{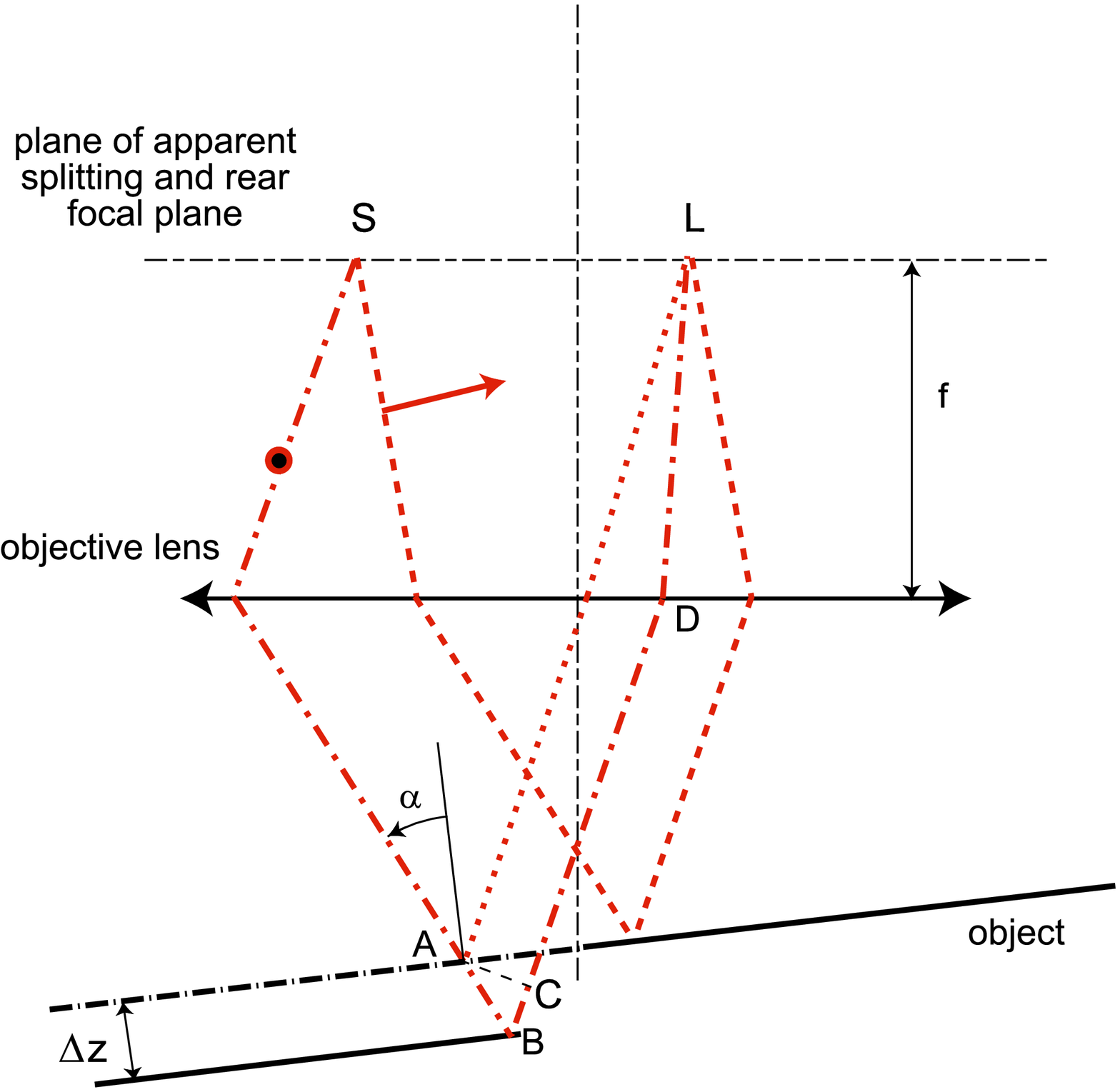}}
       \caption{Ray tracing in the case of a tilted and stepped sample (height $\Delta z$).}
       \label{fig:marche}
\end{figure}
The previous ray tracing is shown in the case of a stepped sample
(height $\Delta z$) in Fig. \ref{fig:marche}. Thanks to the
Fermat's principle, tilting the sample doesn't induce an extra
phase shift in the objective-sample path (\ie regardless of the
phase shift induced by the Wollaston prism). The dot line is for
the ray reflected under the previous conditions. The optical phase
difference $\phi_h$ arising from the step corresponds to a travel
along $[AB]$ and $[BC]$, where $C$ is the orthogonal projection of
$A$ on $[BD]$. In a medium which refractive index is $n$, this
optical phase difference reads
\begin{equation}
\phi_h = \frac{2 \pi}{\lambda} n \times ( AB + BC ) = \frac{4
\pi}{\lambda} n  \Delta z \cos(\alpha)
\end{equation}
where $\alpha$ is the incidence angle on the object. If the
numerical aperture of the objective is low enough, $\alpha$
remains small and the above expression expands
\begin{equation}\label{eq:approx_normale}
\phi_h \simeq \frac{4 \pi n \Delta z}{\lambda}
\end{equation}
As the numerical aperture of the objective increases, the previous
expansion is no longer valid and the equation
(\ref{eq:approx_normale}) is replaced, for numerical apertures
less than $0.3$, by
\begin{equation}\label{eq:phih}
\phi_h=\frac{4 \pi n \Delta z}{\iota \lambda}
\end{equation}
where $\iota$ is a scale parameter depending on the numerical
aperture (see appendix \ref{annexe:on}).

\subsubsection{Optical phase arising from the local surface
orientation}\label{ss:phi_wol_pente}

In section D\ref{sss:phase_marche}, the two orthogonally polarized
rays cross on the PAS. This is no longer true if the two rays
experience a different surface orientation.
\begin{figure}[htb]
       \centerline{\epsfxsize=.5\hsize \epsffile{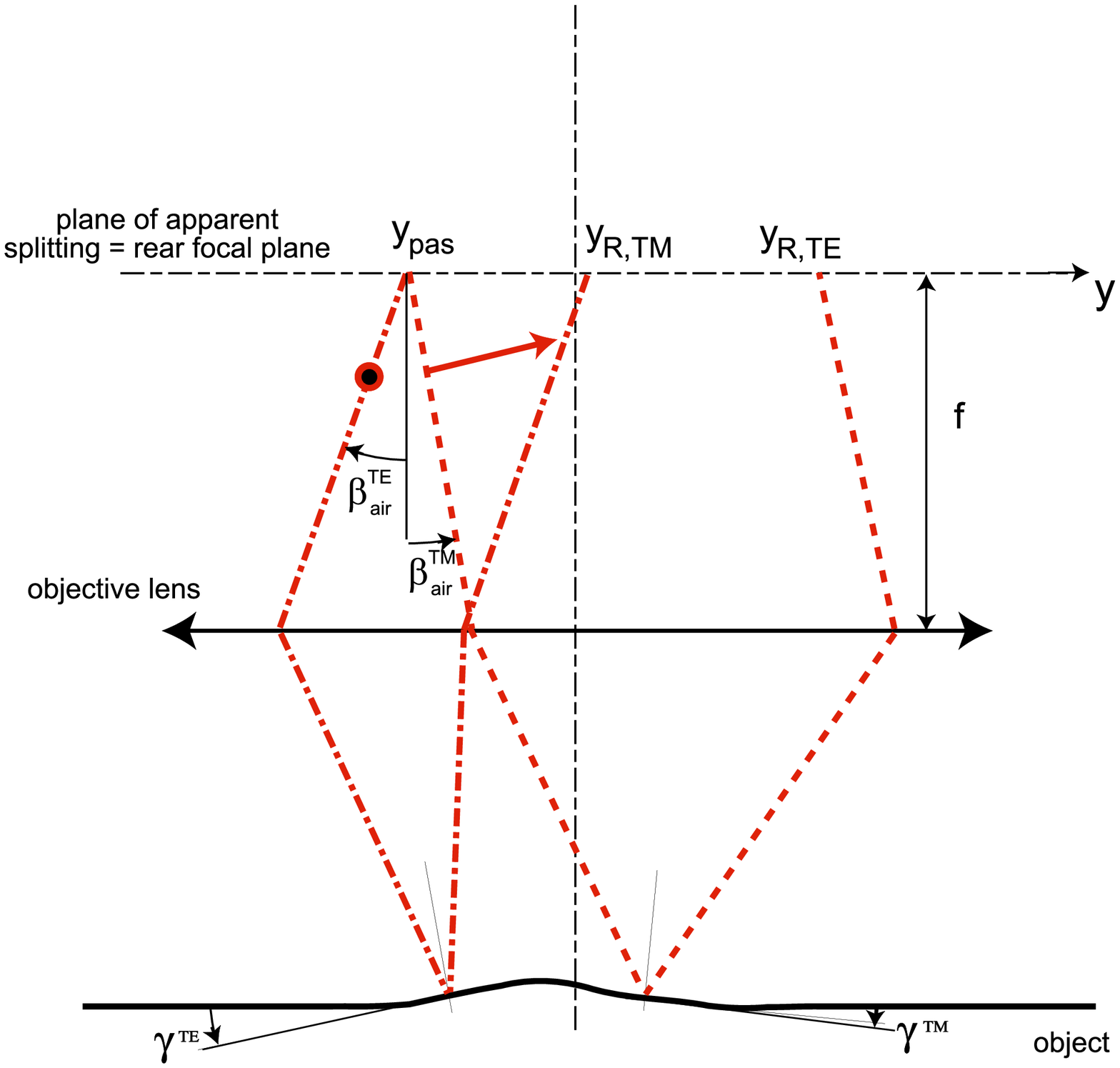}}
       \caption{Ray tracing for a sample subjected to slope variations.}
       \label{fig:focal_wollaston_pente}
\end{figure}
Let us denote by $\gamma^{TE}$ (resp. $\gamma^{TM}$) the local
surface orientation experienced by the $TE$ (resp. $TM$) ray. For
a ray emerging from the PAS at the position $y$ (in the prism's
frame), the reflected rays cross the PAS at the abscissa (figure
\ref{fig:focal_wollaston_pente})
\begin{eqnarray}
y_{R,TE}=\frac{-y+f \tan(2 \gamma^{TE})}{1+\frac{y}{f} \tan(2 \gamma^{TE})}\\
y_{R,TM}=\frac{-y+f \tan(2 \gamma^{TM})}{1+\frac{y}{f} \tan(2
\gamma^{TM})}
\end{eqnarray}
Where $f$ is the focal length of the objective lens. The optical
path travelled by the two rays reads (with $\alpha_1=0$)
\begin{eqnarray}
l_{ar}^{TE}&\simeq& l^{TE}(T_w+y,0) + l^{TE}(T_w+y_{R,TE},0)\\
l_{ar}^{TM}&\simeq & l^{TM}(T_w+y,0) + l^{TM}(T_w+y_{R,TM},0)
\end{eqnarray}
and the optical phase difference reads (to the first order with
respect to the surface orientation)
\begin{equation}
\phi_{W}=\frac{2 \pi}{\lambda}(l_{ar}^{TE}-l_{ar}^{TM}) = \frac{4
\pi}{\lambda}(n_E-n_o)\tan(\theta)
  \left(1+ \left(\frac{y}{f}\right)^2 \right)
f(\gamma_{TE}+\gamma_{TM})
\end{equation}
One should highlight that this optical phase difference doesn't
arise from the sample itself, but from the optical phase imbalance
it introduces when the rays travel back through the Wollaston
prism. Moreover, this phase difference depends on the abscissa
$y$, and then on the angle $\alpha$. It's also worth noting that a
global tilt of the specimen (\ie $\gamma_{TE}=\gamma_{TM} \neq 0$)
induces an extra optical phase. It arises from the previous
remarks that the relation between the local surface orientations
and the induced optical phase depends on the prism-objective
group. This relation may be rewritten
\begin{equation}\label{eq:phiw}
\phi_{W}= \phi_{ori}+\phi_{tilt} = \frac{\partial \phi_W}{\partial
\gamma} (\gamma_{TE}-\gamma_{TM}) + \phi_{tilt}
\end{equation}
with
\begin{equation}
\phi_{tilt}= 2 \frac{\partial \phi_W}{\partial \gamma} \gamma_{TM}
\end{equation}
The scalar $\frac{\partial \phi_W}{\partial \gamma}$ then
describes the prism-objective group. It may be obtained when
measuring the optical phase when tilting a reasonably flat mirror.
\begin{figure}[htb]
       \centerline{\epsfxsize=.5\hsize \epsffile{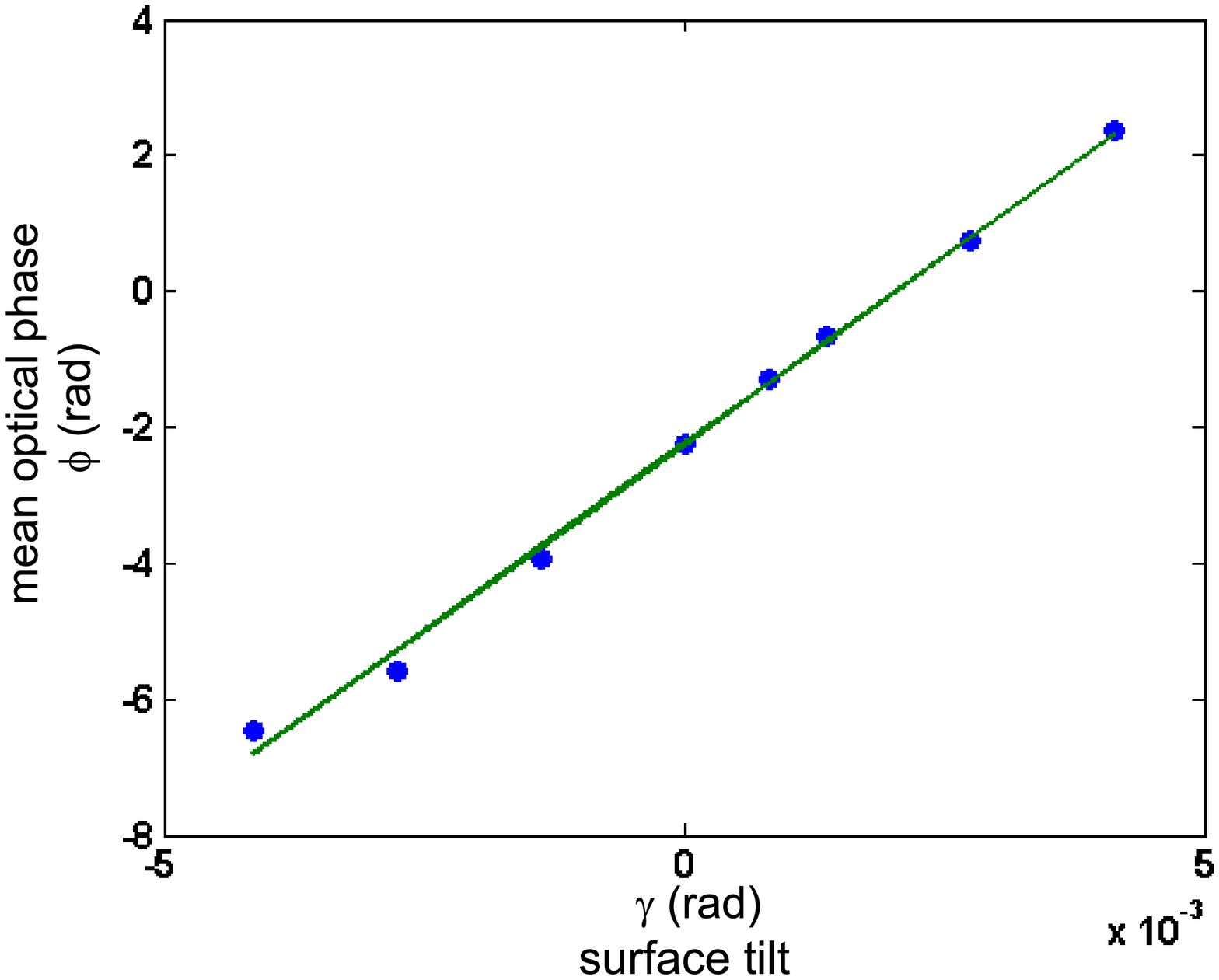}}
       \caption{Calibration measurement of the mean phase induced by a sample as a function of its tilt.}
       \label{fig:etalonnage_ori}
\end{figure}
The figure \ref{fig:etalonnage_ori} shows the result of such a
calibration for a quartz Wollaston prism ($\theta \simeq
18^\circ$) and a water immersion objective lens (focal length $18
mm$,  NA $0.3$, used to obtain the interference pattern in
Fig.\ref{fig:ima_interf}). The relation is linear, and the fitted
slope is
\begin{equation}
2 \frac{\partial \phi_W}{\partial \gamma} = 1.1 \times 10^3
\end{equation}
This simple experiment allows one to obtain the sensibility
$\frac{\partial \phi_W}{\partial \gamma}$ to a local orientation
gap by tilting the whole sample. \par Finally, the total measured
optical phase difference $\phi$ may be written as the following
sum
\begin{equation}\label{eq:phi_fin}
\phi=\phi_{W0}(T_w)+\phi_{PAS}(\delta_{PAS})+\phi_h(\Delta
z)+\phi_{W}(\gamma_{TE},\gamma_{TM})
\end{equation}
where the two first terms (related to $\phi_0$) account for the
path difference introduced by the Wollaston prism itself : an
homogeneous term $\phi_{W0}$ and a linear term along the shear
direction $\phi_{PAS}$ which vanishes when the PAS matches the
rear focal plane of the objective lens. The two last field terms
(related to $\phi_m$) originate from the surface topography :
height variations $\phi_h$ and slope field, $\phi_{W}$ including
the contribution of the mean tilt of the sample $\phi_{tilt}$ as
well as slope variations $\phi_{ori}$.

\section{Phase integration, retrieval and unwrapping}
\subsection{Principle}

The optical flux collected by the pixel indexed by $(l,m)$ on the
CCD matrix can be formally written according to
(\ref{eq:form_inter}) as
\begin{equation}\label{eq:form_interf_mod}
I(l,m,t)=I_{0}+A\cos [\phi (l,m)+\psi_{mod} (t)],
\end{equation}
where $\phi(l,m)$, which is to be determined, is the optical phase
introduced by the sample and the Wollaston prism. The phase
modulation introduced by the photoelastic modulator reads
\begin{equation}
\psi_{mod} (t)=\psi_{0}\sin (2\pi f_{mod} t+\theta_{mod} ).
\end{equation}
The angles $\psi _{0}$ et $\theta_{mod}$ are two parameters that
can be chosen among many couples. The algorithm to obtain $\phi$
uses four integrating buckets \cite{dubois_4f}: if~$T~=~1/f_{mod}$
is the modulation period, four images of the interference pattern
can be captured during the period~$T$, so that each image results
from the integration of the optical flux during a quarter of one
period. One obtains four images $ E_{p} $, for $ p=1,2,3,4 $

\begin{equation}
E_{p}=\int _{\frac{(p-1)T}{4}}^{\frac{pT}{4}}I(t)\mathrm{d}t
\end{equation}

Combining with Eq. (\ref{eq:form_interf_mod}) allows one to obtain
\begin{eqnarray}\label{eq:ep}
E_p&=&\frac{T}{4}(I_0+A J_0(\psi)\cos(\phi)) \nonumber \\
& + & \frac{T A
\cos(\phi)}{\pi}\sum_{n=1}^{\infty}\frac{J_{2n}(\psi_0)}{2n}
\left[ \sin(n
p \pi + 2 n \theta_{mod})-\sin(n (p-1) \pi + 2 n \theta_{mod}) \right] \nonumber \\
& - & \frac{T A
\sin(\phi)}{\pi}\sum_{n=0}^{\infty}\frac{J_{2n+1}(\psi_0)}{2n+1} \nonumber \\
&& \times \left(\cos(\frac{\pi}{2}(2n+1)(p-1)+(2n+1)\theta_{mod})
\right. \nonumber \\
&& \hspace{1cm} \left.
-\cos(\frac{\pi}{2}(2n+1)p+(2n+1)\theta_{mod}) \right)
\end{eqnarray}
where $J_n$ is the first kind Bessel function of $n$ order. The
images $E_p$ depend on $I_0$, $\cos(\phi)$ and $\sin(\phi)$, so
that using four independent images provides enough information to
recover $\phi$. Classical algorithms \cite{dubois_4f} use
particular linear combinations
\begin{equation}\label{eq:def_sigma}
\begin{array}{lll}
\Sigma_s &=& -(E_{1}-E_{2}-E_{3}+E_{4})=\frac{4TA}{\pi}\Gamma_s
\sin(\phi) \\
\Sigma_c &=& -(E_{1}-E_{2}+E_{3}-E_{4})=\frac{4TA}{\pi}\Gamma_c
\cos(\phi)
\end{array}
\end{equation}
with
\begin{equation}\label{eq:def_gamma}
\begin{array}{lll}
\Gamma_s&=&\sum_{n=0}^{\infty}(-1)^n \frac{J_{2n+1}(\psi_0)}{2n+1}
\sin \left[ (2n+1) \theta_{mod} \right] \\
\Gamma_c&=&\sum_{n=0}^{\infty} \frac{J_{4n+2}(\psi_0)}{2n+1}
\sin\left[ 2(2n+1) \theta_{mod} \right]
\end{array}
\end{equation}
The optical phase is then usually recovered by using an ``atan2''
function with the arguments provided by the equations
(\ref{eq:def_sigma}) with a $(\psi_0,\theta_{mod})$ couple
satisfying $\Gamma_s=\Gamma_c$. The indicator
\begin{equation}\label{eq:def_indicator}
\Upsilon^2=\Sigma_c^2+\Sigma_s^2
\end{equation}
is then independent of the phase $\phi$ and defines the intensity
of the signal experiencing the phase $\phi$. As a consequence, the
phase obtained from the previous algorithm is reliable provided
that $\Upsilon^2$ is high enough, \ie if the fringe contrast
$\frac{A}{I_0}$ is sufficient. This condition may be difficult to
satisfy when the topography is described by a large number of
closely packed interference fringes (see
Fig.\ref{fig:ima_interf}). It's worth noting that the set of four
images provides an over-determinated set of data. This redundancy
is then exploited to provide a more reliable phase measurement.

\subsection{Least-square phase retrieval}
For each pixel of the CCD array, the set of Eq. (\ref{eq:ep}) may
we rewritten as a linear system of equations
\begin{equation}
\TT{M} \T{P} = \T{E}
\end{equation}
where the parameters vector $\T{P}$ reads ($(\cdot)^t$ denotes the
transpose of $(\cdot)$)
\begin{equation}
\T{P}^t=\left[ \frac{T I_0}{4}, \frac{T A}{\pi} \cos(\phi),
\frac{T A}{\pi} \sin(\phi) \right]
\end{equation}
and the images vector
\begin{equation}
\T{E}^t=\left[ E_1, E_2, E_3, E_4 \right]
\end{equation}
The matrix $\TT{M}$ is built from the modulation parameters
\begin{equation}
\TT{M}= \left[ {\begin{array}{*{20}c} 1 & c(1,\psi_0,\theta_{mod})
 & s(1,\psi_0,\theta_{mod}) \\
1 & c(2,\psi_0,\theta_{mod})
 & s(2,\psi_0,\theta_{mod}) \\
1 & c(3,\psi_0,\theta_{mod})
 & s(3,\psi_0,\theta_{mod}) \\
1 & c(4,\psi_0,\theta_{mod})
 & s(4,\psi_0,\theta_{mod}) \\
\end{array}}
\right]
\end{equation}
with
\begin{eqnarray}
c(p,\psi_0,\theta_{mod}) &=&
\sum_{n=1}^{\infty}\frac{J_{2n}(\psi_0)}{2n} \left[ \sin(n p
\pi + 2 n \theta_{mod})-\sin(n (p-1) \pi + 2 n \theta_{mod}) \right] \\
s(p,\psi_0,\theta_{mod}) &=&
-\sum_{n=0}^{\infty}\frac{J_{2n+1}(\psi_0)}{2n+1} \nonumber \\
&& \times \left(\cos(\frac{\pi}{2}(2n+1)(p-1)+(2n+1)\theta_{mod})
\right. \nonumber \\
&& \hspace{1cm} \left.
-\cos(\frac{\pi}{2}(2n+1)p+(2n+1)\theta_{mod}) \right)
\end{eqnarray}
The matrix $\TT{M}$ is then independent of the considered point.
For each pixel of the CCD array, the four images describe the
vector $\T{E}$, and the solution parameters vector $ \T{P}_{sol} $
is obtained as a minimizer of
\begin{equation}
\eta^2(\T{P})=\left( \TT{M} \T{P} - \T{E}   \right)^t \left(
\TT{M} \T{P} - \T{E}   \right)
\end{equation}
and is the solution of the square linear system
\begin{equation}
\TT{M}^t \TT{M}  \T{P}_{sol}=\TT{M}^t \T{E}
\end{equation}
The couple $\left( \frac{T A \cos(\phi_{sol})}{\pi}, \frac{T A
\sin(\phi_{sol})}{\pi} \right)$ is then extracted, and used as the
argument of a standard ``atan2'' function to provide a less
corrupted value of the phase.
\begin{figure}[htb]
       \centerline{\epsfxsize=.5\hsize \epsffile{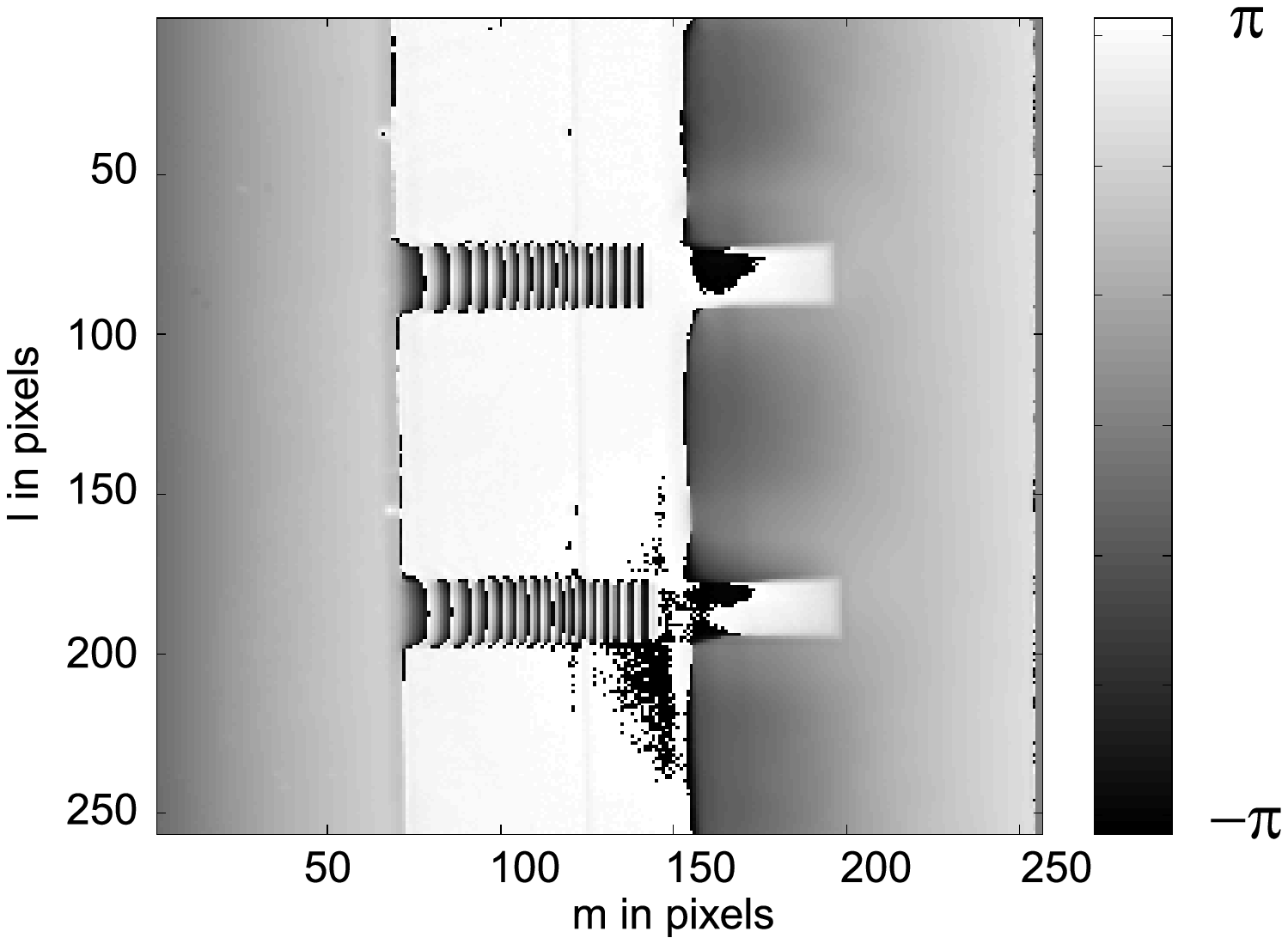}}
       \caption{Typical wrapped phase map obtained in water with two $70 \times 20 \times 0.84 \mu m^3$ microcantilevers and a shear distance $d = 53.4 \mu m$ (NA=0.3).}
       \label{fig:ima_phi}
\end{figure}
The figure \ref{fig:ima_phi} shows the phase map obtained from the
scene presented in Fig. \ref{fig:ima_interf} using the
least-square algorithm.

\subsection{Phase unwrapping} \label{ss:phiunw}
The use of the phase integration technique also allows one to
reconsider the phase unwrapping problem. In the previous section,
the information derived from the experiments is a couple $(X,Y)$
proportional to $(\cos(\phi),\sin(\phi))$
\begin{equation}
(X,Y)=C (\cos(\phi),\sin(\phi))
\end{equation}
where $C=\frac{4TA}{\pi}\Gamma_{s,c}$ when the linear combinations
(\ref{eq:def_sigma}) are used, and $C=\frac{T A}{\pi}$ when the
least-square algorithm is used. In the first case, the couples
$(X,Y)$ and $(\Sigma_c,\Sigma_s)$ are equal. For a sake of
simplicity, let us consider that a wrapped value of the phase
$\phi_p$ is then obtained by using
\begin{equation}\label{eq:atan2}
\phi_p=atan2(Y,X)
\end{equation}
$\phi_p$ lies then in $[-\pi,\pi]$. Recovering the unwrapped value
of the phase may turn into a brain-racking task when the phase-map
is poorly discretized and the phase jumps are densely packed.
Numerous algorithms are available in the literature, based either
on local phase unwrapping \cite{cusack} or on global phase
unwrapping \cite{he}. Contrary to most of these algorithms, which
use a wrapped phase map as an input, we propose here to use the
two fields $X$ and $Y$.

\begin{figure}[htb]
       \centerline{\epsfxsize=.5\hsize \epsffile{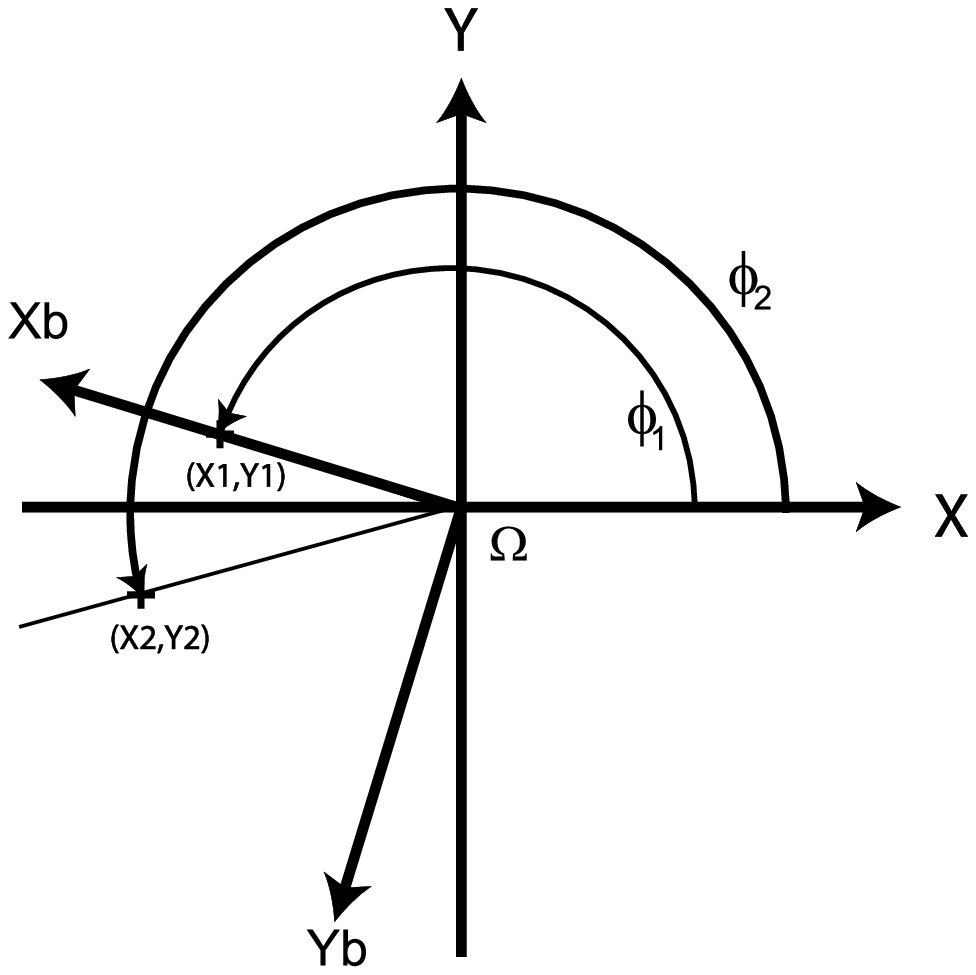}}
  \caption{Phase unwrapping principle.}
  \label{fig:phi_unwrap}
\end{figure}

The figure \ref{fig:phi_unwrap} shows a typical case of phase jump
between two adjacent points 1 and 2. Their true phase $\phi_1$ and
$\phi_2$ is under scrutiny. As $X_1 \simeq X_2 \neq 0$ and $Y_1
Y_2<0$, using the ``atan2'' function induces a phase jump when
moving from 1 ($\phi_{p,1} \simeq \pi$) to 2 ($\phi_{p,2} \simeq
-\pi$). Let us then define the frame
$(\overrightarrow{Xb},\overrightarrow{Yb})$, obtained by rotating
the $(\overrightarrow{X},\overrightarrow{Y})$ frame until the
point 1 $(X_1,Y_1)$ belongs to the $(\Omega,\overrightarrow{Xb})$
axis. The phase $\phi_2-\phi_1$ of the point 2 in the frame
$(\overrightarrow{Xb},\overrightarrow{Yb})$ is then obtained
according to
\begin{equation}\label{eq:phi_un_diff}
    \tan[\phi_2-\phi_1]=
    \frac{-X_2 sin(\phi_1) + Y_2 cos(\phi_1)}{X_2 cos(\phi_1) + Y_2 sin(\phi_1)}
\end{equation}
providing the gap between the \emph{true} phase values $\phi_1$
and $\phi_2$, thus defining (if 1 and 2 are for adjacent pixels),
the true phase gradient, modulo $2 \pi$. The couple $(X_2,Y_2)$ is
obtained of the four images using one of the two algorithms
previously described. $cos(\phi_1)$ and $sin(\phi_1)$ are deduced
from the wrapped value $\phi_{p,1}$. The phase gradient may then
be integrated to provide a true phase map.
\begin{figure}[htb]
       \centerline{\epsfxsize=.5\hsize \epsffile{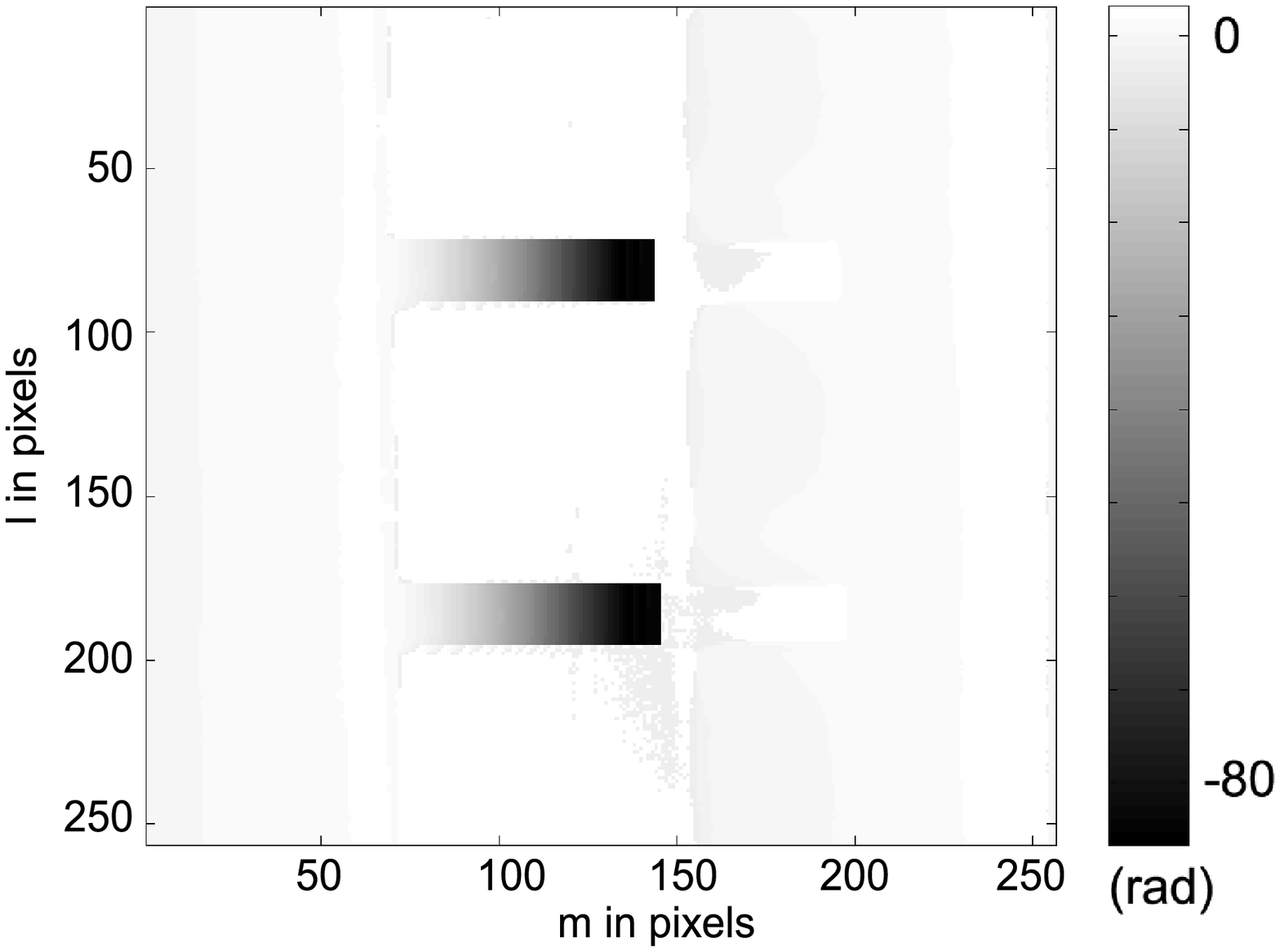}}
       \caption{Typical unwrapped phase map obtained in water with two $70 \times 20 \times 0.84 \mu m^3$ microcantilevers and a shear distance $d=53.4 \mu m$ (NA=0.3).}
       \label{fig:ima_phi_un}
\end{figure}
The figure \ref{fig:ima_phi_un} shows the phase map obtained from
the wrapped phase map shown in Fig.\ref{fig:ima_phi} using the
described phase unwrapping technique.

\subsection{Reproducibility of the phase measurement}
The phase measurement reproducibility is assessed by measuring
twice the phase map arising from the same differential topography
of a reflective object. One gets two phase fields $\phi_-(l,m)$
and $\phi_+(l,m)$. Assuming that each of these fields is the sum
of deterministic part $\phi_d(l,m)$ and of a random part
\begin{eqnarray}
\phi_-(l,m)=\phi_d(l,m)+b_-(l,m) \\
\phi_+(l,m)=\phi_d(l,m)+b_+(l,m)
\end{eqnarray}
the difference between the two phase fields provides then a
realization of the difference between two realizations of the
noise $b(l,m)$
\begin{equation}
(\phi_+ - \phi_-)(l,m)=(b_+ - b_-)(l,m)
\end{equation}
\begin{figure}[htb]
       \centerline{\epsfxsize=.5\hsize \epsffile{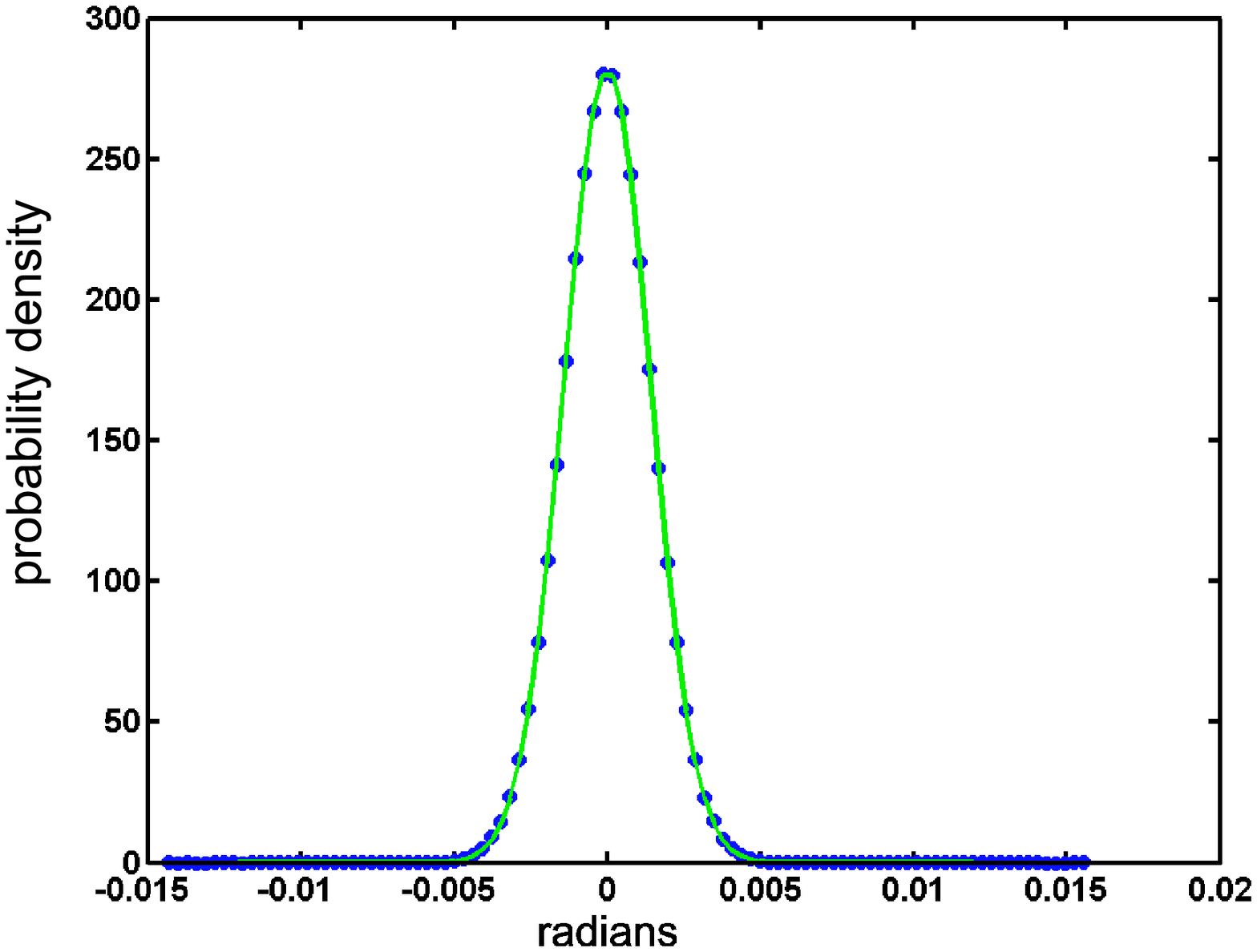}}
       \caption{Typical experimental phase noise probability density (dots) and its fit by a Gaussian distribution (solid line).}
       \label{fig:distrib_phase}
\end{figure}
The figure \ref{fig:distrib_phase} shows a typical probability
density of the variable $(\phi_+ - \phi_-)$ obtained with the
set-up described in Section \ref{ss:nom}. The dots are the
experimental distribution, whereas the solid line is a
least-square fit using a Gaussian, zero-mean, distribution. The
agreement is excellent, and allow us to consider $(\phi_+ -
\phi_-)$ as a random real number of variance $2 \sigma^2$, if $b$
is a random variable which variance is $\sigma^2$.
\begin{figure}[htb]
       \centerline{\epsfxsize=.5\hsize \epsffile{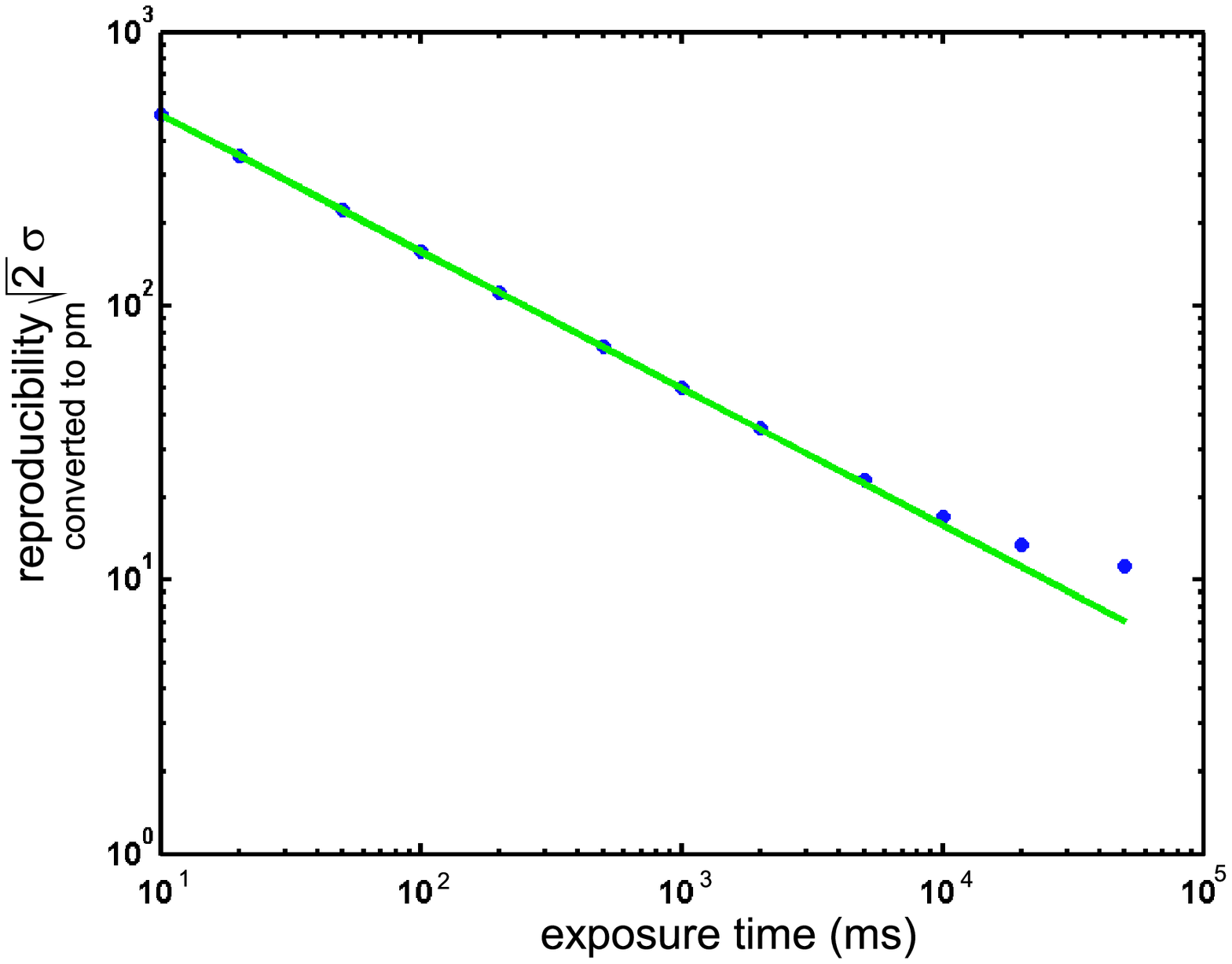}}
       \caption{Estimation of the reproducibility on the measurement of a differential topography as a function of the exposure time.}
       \label{fig:reprod}
\end{figure}
The figure \ref{fig:reprod} shows the evolution of $\sqrt{2
\sigma^2}$ (converted to heights variation assuming
(\ref{eq:approx_normale})) as a function of the exposure time
$t_{exp}$ used to form each intensity image. The experimental
variance grows as $t_{exp}^{-\frac{1}{2}}$, until the exposure
time reaches several tens of seconds. The reproducibility is
therefore controlled by the exposure time, and the achieved level
is (converted to height variations) close to 10 pm.
\section{Recovering the displacement field}
\subsection{Principle} \label{ss:depl_pr}
The calculations presented in sections 2\ref{ss:phi_wollaston} and
2\ref{ss:phi_topo} allow one to derive an expression for the
measured phase $\phi$ as a function of the surface topography.
Assuming that the pixel size (in the object plane) is small enough
comparing to the object's size, one proposes for a sake of
simplicity to formulate the inversion problem in the object's
plane. Inserting Eq. (\ref{eq:phih}) and Eq. (\ref{eq:phiw}) into
Eq. (\ref{eq:phi_fin}) allows one to write the total measured
phase difference at the point $(x,y)$
\begin{equation}\label{eq:def_deltaphi}
\phi(x,y)=\phi_0(x,y)+\phi_{tilt}+\Phi(x,y)
\end{equation}
Assuming that the shear direction is parallel to the $y$ axis, and
that the local surface orientation $\gamma$ in Eq. (\ref{eq:phiw})
is given by the first derivative of the topography, the relation
between the measured information $\Phi$ and the topography reads,
for a shear distance $d$
\begin{equation}\label{eq:def_phi}
\Phi(x,y)=\frac{4 \pi n}{\iota \lambda}
\left(z(x,y+\frac{d}{2})-z(x,y-\frac{d}{2}) \right)
+\frac{\partial \phi_W}{\partial \gamma} \left(\frac{\partial
z}{\partial y}(x,y+\frac{d}{2})-\frac{\partial z}{\partial
y}(x,y-\frac{d}{2}) \right)
\end{equation}
where $z(x,y)$ is the topography. To discriminate between the
phase arising from the slope variations and those arising from
height variations, one proposes to expand the topography on a
functions basis
\begin{equation}\label{eq:decomposition_z}
z(x,y)=\sum_s \mu_s z_s(x,y)
\end{equation}
so that the reference (initial) measured phase field reads
\begin{equation}
\phi_{ref}(x,y) = \phi_0(x,y)+\phi_{tilt}+ \sum_s \mu_s
\Phi_s(x,y)
\end{equation}
One should highlight that the shearing amount $d$ is here
introduced explicitly to compute the functions basis $\Phi_s(x,y)$
by inserting Eq.(\ref{eq:decomposition_z}) into
Eq.(\ref{eq:def_phi}). If the topography is subjected to an
out-of-plane displacement field $w(x)$, $\phi_{ref}$ is changed
into $\phi_{w}$ and the new topography is described by
\begin{eqnarray}
z(x,y)+w(x,y)=\sum_s (\mu_s+\nu_s) z_s(x,y)
\end{eqnarray}
so that the displacement field is also expanded on the same
functions basis
\begin{eqnarray}
w(x,y)=\sum_s \nu_s z_s(x,y)
\end{eqnarray}
The knowledge of the numerical coefficients involved in the
definition (\ref{eq:def_phi}) of $\Phi(x,y)$ combined with the
choice of a functions basis allows one to recover the displacement
field by computing the $\nu_s$ coefficients minimizing
\begin{equation}
\eta_\nu^2(\nu)=\int{\left( \sum_s \nu_s \Phi_s(x,y) - (\phi_{w} -
\phi_{ref})(x,y) \right)^2 dx dy}
\end{equation}
that is solving the linear system
\begin{equation}
\TT{N} \T{\nu}=\T{F}
\end{equation}
with
\begin{eqnarray}
N_{si} &=&  \int{ \Phi_s(x,y) \Phi_i(x,y) dx dy}\\
F_s &=& \int{ \Phi_s(x,y) (\phi_{w} - \phi_{ref})  dx dy}
\end{eqnarray}

\subsection{Example}
The micro-cantilevers shown on Fig.\ref{fig:ima_interf} are placed
into a fluid cell, filled with milliQ water which temperature is
controlled thanks to a feedback loop controlled Peltier device.
These cantilevers are multi-layer cantilevers, made of a silica
layer (770 nm), a titanium layer (20 nm) and a gold layer (50 nm).
These cantilevers are then subjected to a bimaterial effect, since
the coefficient of thermal expansion of the gold layer is almost
ten the one of the silica layer. The optical set-up is also used
with the Wollaston prism-objective couple calibrated in Section
\ref{ss:phi_wol_pente}. A reference phase map of the cantilever is
captured at $22.6 \degre C$. The temperature is then increased to
reach $24.1 \degre C$, and a new phase map is captured.
\begin{figure}[htb]
       \centerline{\epsfxsize=.5\hsize \epsffile{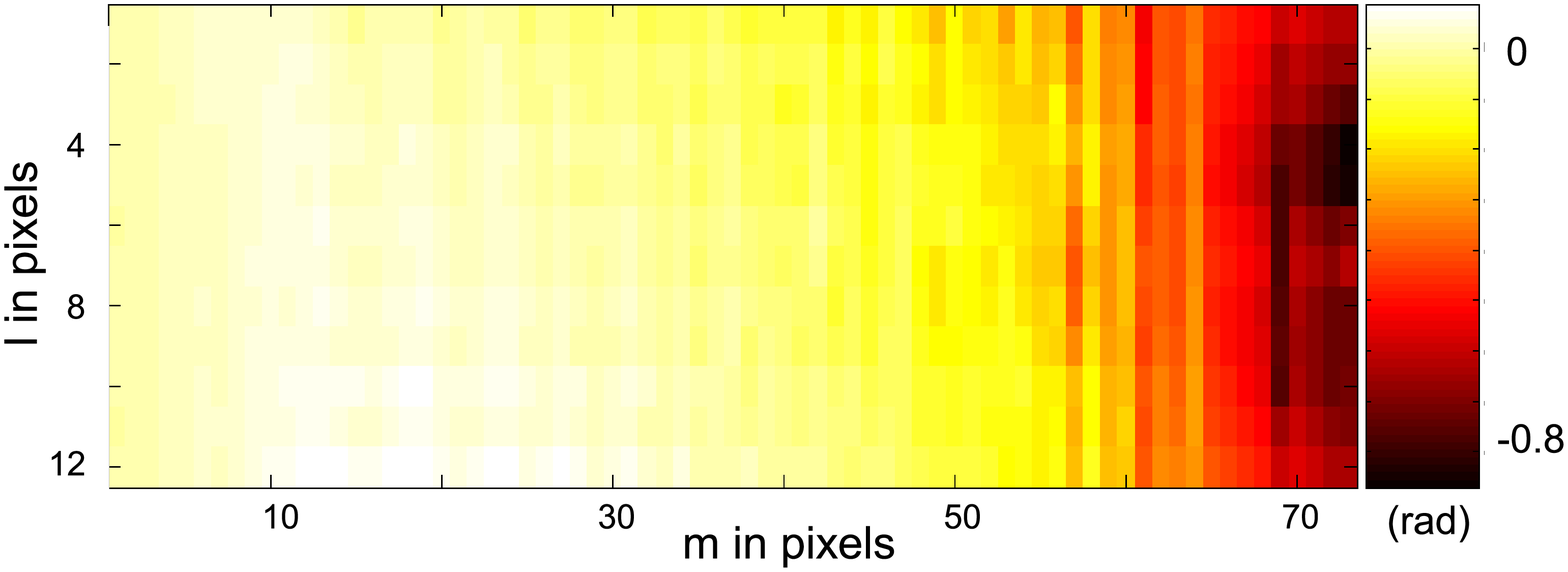}}
       \caption{Measured phase map change when the cantilever is subjected to bimaterial effect.}
       \label{fig:champ_phase}
\end{figure}
The phase gradients are computed according to the algorithm
described in Section 3\ref{ss:phiunw}. Recovering the
two-dimensional phase map from the phase gradients is an
over-constrained problem, so that it is possible to avoid some
unreliable data. Reliable data are then located where the
indicator defined by Eq.(\ref{eq:def_indicator}) is greater than a
user-defined threshold. The figure \ref{fig:champ_phase} shows the
measured phase map change when the cantilever is submitted to
bimaterial effect. One distinguish the substrate, which is not
subjected to any modification. One then remark the phase change is
not homogeneous across the cantilever.

\begin{figure}[htb]
       \centerline{\epsfxsize=.5\hsize \epsffile{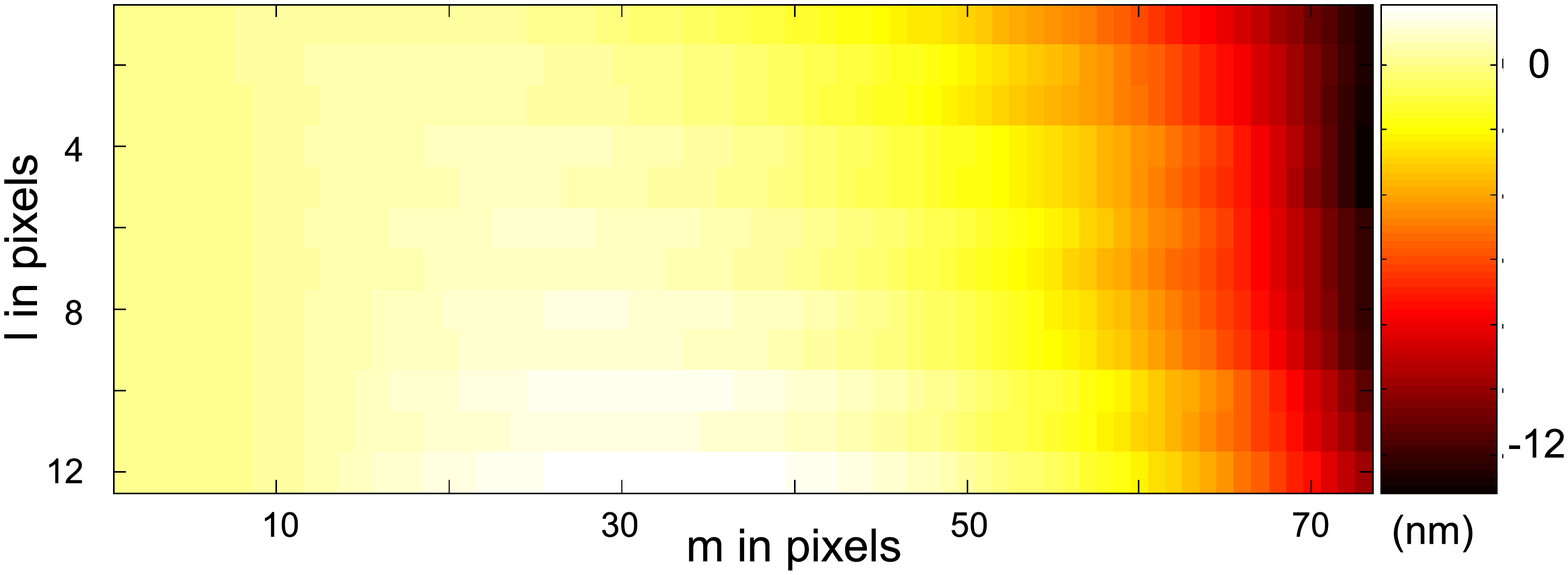}}
       \caption{Displacement field calculated from the measured phase map change shown in Fig.\ref{fig:champ_phase}.}
       \label{fig:champ_depl}
\end{figure}

The displacement field is then recovered using the algorithm
described in Section 4\ref{ss:depl_pr}. The function basis is
chosen to be able to represent the expected mechanical effects. As
a consequence, one chooses cubic hermite polynomials
\cite{timoshenko} by part along the $y$ direction. In the present
example, four elements along the cantilever are found to be
sufficient to describe heterogeneous effects, and the projection
is made independently for each line across the beam. The figure
\ref{fig:champ_depl} shows the resulting displacement field, which
clearly exhibits an heterogeneous behavior across the cantilever.
One should emphasize, that the free end of the cantilever
experiences a phase change of 0.8 rad (Fig.\ref{fig:champ_phase}).
This value is to be compared to the one arising only from a 12 nm
height modification (Fig. \ref{fig:champ_depl}), which is 0.18 rad
(first term in Eq.(\ref{eq:def_phi})) ; thus demonstrating the
necessity of taking both height and slope changes into account.
Moreover, the displacement field shows the cantilever part close
to the anchoring rises up as the temperature increases, whereas
the cantilever as a whole bends down. This is thought to be the
signature of an under-etched cantilever, as the remaining silica
pedestal dilates and pushes the surface up. This example typically
illustrates the extensive amount of informations provided by
full-field measurement compared to pointwise ones.

\section{Conclusions}

\par A Nomarski imaging interferometer is used to measure the
differential topography of reflective objects. A phase modulation
is introduced to measure the phase map arising from the object
with a sensitivity independent on the actual phase value. The
phase measurement is shown to be shot-noise limited, and a
measurement reproducibility of almost 10 pm is achieved. The use
of four integrating buckets with a least-square algorithm is shown
to improve the phase retrieval in case of low fringe constrast.
Moreover, this allows to reconsider the phase unwrapping problem,
so that it is easy to deal with highly curved objects.

\par The physical origin of the measured optical phase is
calculated, and exhibits that the phase difference is due to both
height and slope variations. A simple calibration procedure allows
one to check for the phase sensitivity to slope variations. These
two effects are then decoupled expanding the displacement field
onto a chosen functions basis.

\par The out-of-plane displacement field of a MEMS cantilever
subjected to bimaterial effect is then recovered thereby proving
the ability of such a set-up to provide a reliable full-field
kinematic measurement without surface modification. This tool
provides then a full displacement field using a common-path
interferometer. Moreover, its stability, as well as its ability to
operate through a wide range of media make this set-up a very
powerful tool for the study of the mechanical behavior of
micro-electro-mechanical systems, providing an extensive amount of
reliable information.

\newpage

\appendix

\section{Appendix : Effect of the numerical aperture}
\label{annexe:on} The section 2\ref{ss:phi_topo} is devoted to the
calculation of the optical phase difference arising from the
topography of the object under scope. This difference depends on
the incidence of the ray on the surface. When using an imaging
system, this difference depends therefore on the numerical
aperture of the objective lens. The expression
(\ref{eq:approx_normale}) was obtained under the assumption that
the incidence angle is low enough. To assess this assumption's
validity, let us consider the contribution of each incidence in
the figure's plane to the total intensity in
Eq.\ref{eq:form_inter}
\begin{equation}
dI(\alpha)= \left( dI_0+dA \cos \left[\frac{4 n \pi \Delta
z}{\lambda} \cos(\alpha) + \psi\right] \right) d \alpha
\end{equation}
where $\psi$ stands for the optical phase independent from the
topography. As the fringe spacing depends on the angle $\alpha$
between the ray and the optic axis, the optical phase arising from
the step in Fig. \ref{fig:marche} is obtained by weighting and
summing the contributions of each rays impinging on the sample at
a point of the field of view
\begin{equation}
I=I_0+A \frac{2}{\sin^2(\alpha_{max})} \int_0^{\alpha_{max}} \cos
\left[\frac{4 n \pi \Delta z}{\lambda} \cos(\alpha) + \psi\right]
P(\alpha)^2 \sin(\alpha) d \alpha
\end{equation}
$\alpha_{max}$ is related to the numerical aperture of the
objective lens $NA$
\begin{equation}
NA=n \sin(\alpha_{max})
\end{equation}
and to the apodization function $P(\alpha)$. The choice of this
function has been widely discussed
\cite{dubois_aperture,sheppard,schulz} and let us the assume
\begin{equation}
P(\alpha)=(\cos (\alpha))^m
\end{equation}
where $m$ is a parameter used to describe the apodization effect.
The equation (\ref{eq:form_inter}) turns into
\begin{equation}\label{eq:int_mod}
I=I_0+A_m F_{NA,m}(\Delta z,\psi)
\end{equation}
for $m=0$ (Herschel condition),
\begin{equation}
F_{NA,0}(\Delta z,\psi) \simeq \frac{\sin(k \Delta
z(1-\cos(\alpha_{max})))}{k \Delta z(1-\cos(\alpha_{max}))} \cos(k
\Delta z(1+\cos(\alpha_{max}))+\psi)
\end{equation}
where $k=\frac{2 n \pi}{\lambda}$. If $m \neq 0$, one can derive
the Taylor expansion of (\ref{eq:int_mod}) with respect to
$\alpha_{max}$
\begin{eqnarray}\label{eq:dev_f_on_m}
F_{NA,m}( \Delta z,\psi)&=&
\frac{2}{\sin^2(\alpha_{max})}\int_0^{\alpha_{max}} \cos (2k
\Delta z \cos(\alpha) + \psi) P(\alpha)^{2m} \sin(\alpha) d \alpha \nonumber \\
& \simeq & \cos(2k \Delta z + \psi)+\left(\frac{1-2m}{4}\cos(2k
\Delta z+\psi)+\frac{1}{2}\sin(2k \Delta z+\psi)k \Delta
z \right)\alpha_{max}^2 \nonumber \\
&& +\left(\frac{1+4(m^2-m-k^2 \Delta z^2)}{24} \cos(2k \Delta z +
\psi) \right. \nonumber \\
&& \left. + \frac{k \Delta z (1 - 4m)}{12}\sin(2k \Delta z + \psi)
\right)\alpha_{max}^4+\ldots
\end{eqnarray}
Let us define the fringe spacing $i_{interf}$, as the value of $k
\Delta z$ where the intensity (\ref{eq:int_mod}) reaches its first
local maximum. As the numerical aperture increases, $i_{interf}
\neq \pi$. Let us thus define the gap
\begin{equation}\label{eq:def_epsi}
\epsilon_{interf}=1-\frac{i_{interf}}{\pi}
\end{equation}
Computing $\epsilon_{interf}$, when the numerical aperture ranges
from $0$ to $0.6$, and $m$ ranges from $0$ to $2$ shows that
$\epsilon_{interf}<0$ so that the fringe spacing increases with
the numerical aperture (see equation (\ref{eq:def_epsi})). if the
numerical aperture is less than $0.4$, $\epsilon_{interf}$ no
longer depends significantly on $m$, so that the numerical
aperture value is sufficient to retrieve $\Delta z$ from the
intensity value. The above correction of the equation
(\ref{eq:approx_normale}) is easy as long as $F_{NA,m}(\Delta
z,\psi)$ is pseudo-periodic in a $\Delta z$ range sufficient to
describe the topography. This is assessed by defining the ratio
\begin{equation}
r_i=\frac{i_{10}}{10 i_{interf}}
\end{equation}
where $i_{10}$ is the $k \Delta z$ value for which
$F_{NA,m}(\Delta z,\psi)$ reaches its tenth local maximum If the
numerical aperture is less than $0.3$, $r_i$ is $1$, thereby
proving that $F_{NA,m}(\Delta z,\psi)$ is pseudo-periodic on the
defined range. The correction of Eq. (\ref{eq:approx_normale})
then reads
\begin{equation}
\phi_h=\frac{4 \pi n \Delta z}{\iota \lambda}
\end{equation}
where $\iota$ depends on the numerical aperture and the
apodization function.
\begin{figure}[htb]
       \centerline{\epsfxsize=.5\hsize \epsffile{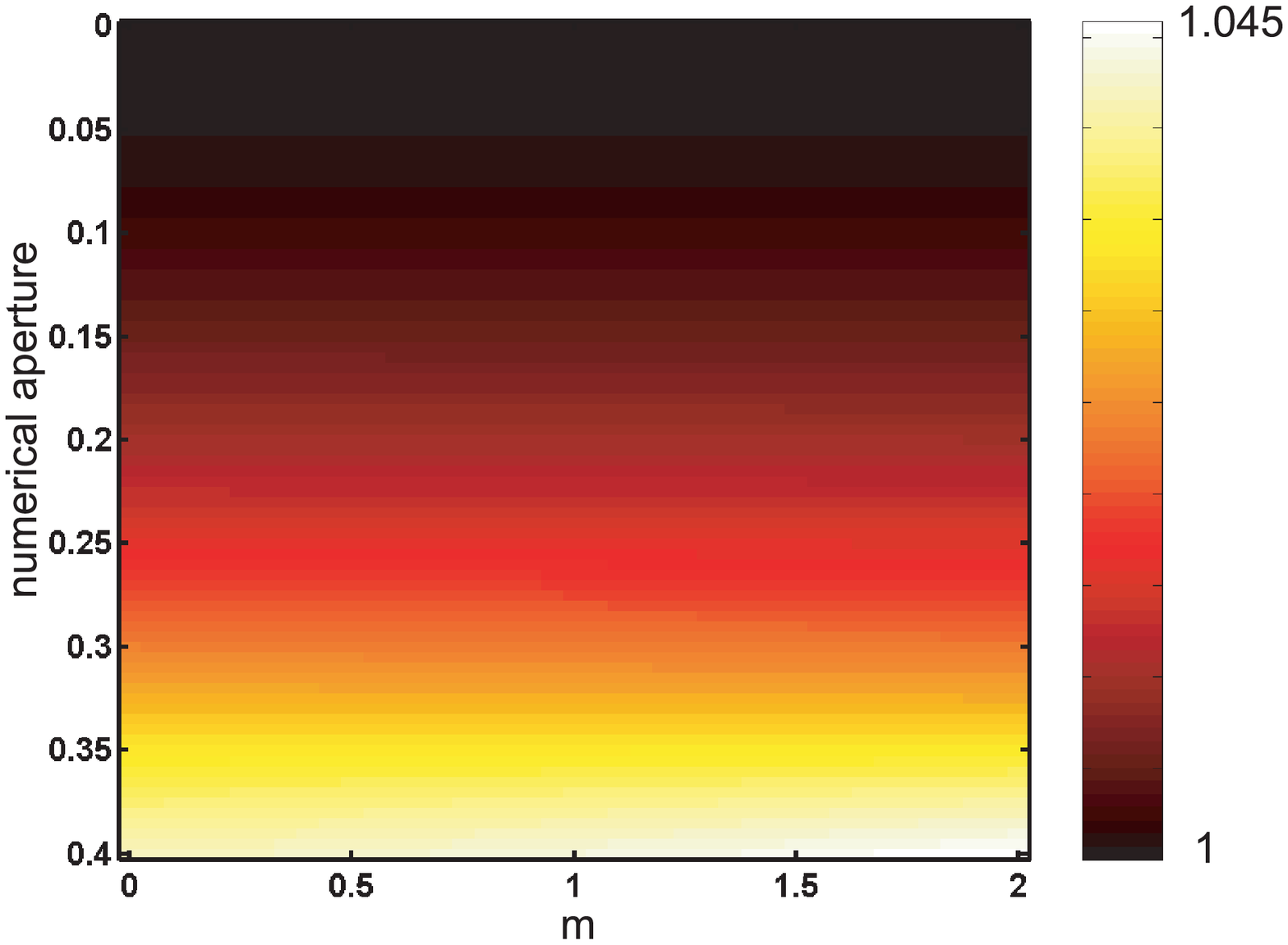}}
       \caption{Evolution of the correction factor $\iota$ when the numerical aperture ranges from $0$ to $0.4$, and the exponent $m$ ranges from $0$ to $2$.}
       \label{fig:coef_an}
\end{figure}
The figure \ref{fig:coef_an} shows the value of $\iota$ as a
function of both the numerical aperture and $m$. If the numerical
aperture is less than $0.3$, $\iota$ no longer depends on $m$, and
therefore allows for a direct retrieval of the phase $\phi_h$.

\newpage

\newpage



\newpage



\end{document}